# Life in Solid Ice?

P. Buford Price
Physics Department, University of California, Berkeley, CA 94720

Some microbes appear to be able to metabolize in glacial ice or permafrost. The rate depends on temperature, nutrient level and bioelement availability, among other factors. I have developed a plausible argument that they do this while confined in veins filled with acidic or saline solution that provides nutrients and elements necessary for growth (78). Here I develop this scenario further and discuss some of its implications for ice-covered planetary bodies and for the origin of life. An accompanying paper (7) discusses plans to test this hypothesis using epifluorescence microscopy of pristine (unmelted) ice samples and an optical biospectrologging tool to assay living and dead microbes in a borehole in glacial ice.

## Inventory of bacteria and archaea in natural environments

Table 1 lists some of the data on concentrations of bacteria and archaea, especially those that live in low-temperature environments. Bacteria are found essentially everywhere on earth in concentrations that do not vary by more than a few orders of magnitude. Archaea are generally nearly as abundant as bacteria when methods for identifying them are used. Table 2 lists some of the types of microbial life found in glacial and permafrost ice.

Challenges to survival include extremes of temperature, pressure, pH, dryness, salinity, oxygen concentration, radiative flux (including sunlight), and availability of nutrients and bio-elements. Glacial ice provides a time machine from which can be extracted microorganisms of known age, whose response to extreme cold and lack of nutrients can be studied. One of the unsolved problems in microbiology is the length of time in conditions of extreme temperatures and nutrient deprivation over which microorganisms can remain viable. No one has yet carried out a thorough study of the depth-dependence of the types of microbes or of the fractions that were viable and metabolizing, that were viable but not culturable, that were dormant, and that were dead. In fact, the definition of "death" may require revision as techniques for cultivating microbes found in ice improve.

We first discuss studies of microbes as a function of depth in the deep Vostok ice core.

Abyzov *et al*. (2) searched for microbes at intervals of ~100 m at depths from 1500 to 2750 m in glacial ice from the Vostok core by passing melted ice through 0.2 µm filters. They reported concentration *vs* depth, studied morphology with SEM and epifluorescence microscopy of stained samples, and studied the fraction of viable organisms in various nutrients. The great majority of the organisms, typically at least 99%, could not be cultivated in the nutrient media used. The concentration of viable organisms decreased with depth, as measured by their consumption of $^{14}$C-labeled compounds. Most of the cells were spore-formers but in a vegetative state. Some were attached to dust particles. The concentration ranged from ~$10^3$ to ~$10^4$ cells/ml and showed a rather good correlation with dust concentration and with climate, suggesting that both dust and microorganisms were deposited in snow preferentially during glacial periods when the wind speed was greatest.

Two groups have published studies of microbes in "accretion ice" located below the glacial ice studied by Abyzov *et al*. (1,2) and ~150 m above Lake Vostok. Karl *et al*. (42) found a concentration of a few hundred cells/ml at 3603 m. In laboratory experiments, some of their microbes developed metabolic activity in the presence of liquid water. Priscu *et al*. (79) detected higher concentrations, ~$10^3$ to ~$3 \times 10^4$ cells/ml, in samples from a depth of 3590 m. Unlike the sample studied by Karl *et al.* (42), his sample contained mineral fragments, which might explain the higher microbial concentration. By comparing microbial concentrations in accretion ice and the water column of Lake Bonney, a permanently ice-covered lake in the McMurdo Dry Valleys, and assuming that the bacterial partitioning was in the same ratio in the Vostok system, Priscu *et al.* (79) estimated that Lake Vostok may contain as many as ~$10^6$ cells/ml.

Table 1. Concentrations of prokaryotes (cells cm$^{-3}$), especially in cold environments

| Location$^{ref.}$ | Bacteria | Archaea |
|---|---|---|
| **Earth's land surface** | | |
| at 0.1 m depth in soil | $10^8$ to $10^{10}$ | ? |
| at 10 m in sediment | $10^7$ to $10^{10}$ | ? |
| at 190-350 m in sediment$^{46,74}$ | ~$10^5$ to $10^6$ | methanogens (Crenarchaeaota) |
| water in rock at 200 m$^{108}$ | 1200 | $2.3 \times 10^5$ |
| at 3 km depth in rock | $3.4 \times 10^5$ | yes |
| **Oceans** | | |
| few m depth$^{43}$ | $3 \times 10^5$ | $4 \times 10^4$ |
| 200 m depth$^{43}$ | $8 \times 10^4$ | $3 \times 10^4$ |
| at 5000 m depth$^{43}$ | 3000 | 4000 |
| sediment at floor$^{71}$ | ~$10^{10}$ | ? |
| sediment 600 m below floor$^{71}$ | $10^6$ | ? |
| ~10,400 m in Pacific trench$^{106}$ | $10^3$ to $10^6$ | ? |
| Antarctic seawater$^{33}$ | ~$7 \times 10^5$ | ~$3 \times 10^{5\ 23}$ |
| **Lakes** | | |
| Alpine$^{72}$ | 1 to $5 \times 10^5$ | up to $3 \times 10^4$ |
| Antarctic (under ice)$^{88,97}$ | up to $4 \times 10^6$ | yes |
| **Clouds**$^{85}$ | 1500 | ? |
| **Snow** | | |
| Sonnenblick Obs.$^{85}$ | 11,000 | ? |
| Ross Ice Shelf$^{33}$ | ~3000 | ? |
| South Pole at -50 $^{11}$ | 200 to 5000 | not seen |
| **Ice cores** | | |
| Vostok (living or dead)$^2$ | 800 to $10^4$ | ? |
| Swiss glaciers, basal ice$^{86}$ | $10^5$ to $6 \times 10^7$ | ? |
| Guliya glacier (China)$^{16}$ | $10^4$ to $5 \times 10^5$ | ? |
| Greenland, Hans Tausen$^{109}$ | ~$3 \times 10^3$ | ? |
| Greenland, GISP2 $^{54}$ | 0.1-0.5 culturable fungi (no other species studied) | ? |
| Basal ice, Ellesmere Island$^{89}$ | >$10^3$ CFU/ml at 4 C | methanogens |
| **Permafrost** | | |
| Siberia$^{107}$ | 300 to $10^8$ | 10-$10^7$ methanogens$^{95}$; $10^5$-$3\times10^7$ denitrifiers$^{81}$ |
| Antarctica$^{31}$ | up to $10^5$ | methanogens |

Further studies, not yet published, by J. Priscu and co-workers indicate that microbial concentrations in Vostok glacial ice are one or two orders of magnitude lower than those reported by Abyzov *et al.* A possible explanation for the difference is that the lower values may result from greater success in eliminating contaminants.

Microbes also have been found in glacial ice from Greenland. Willerslev *et al.* (100) used molecular techniques to identify 57 distinct taxa of eukaryotes from the Hans Tausen ice core at depths corresponding to ~2000 and ~4000 yr. Gruber and Jaenicke (109) estimated concentrations of ~3000/ml in the same core. Ma *et al.* (54) identified fungi with ages as great as 140,000 yr taken from the GISP2 ice core. Culturable fungi alone comprised ~0.1-0.5 cells/ml. They did not measure the total concentration of microbes.

Table 2. Examples of genera detected in glacial and permafrost ice

| Types | Locations [ref.] |
|---|---|
| 23 genera, mostly similar to spore-forming *Bacilli* or *Actinobacteria* | Glacial ice from various locations[16] |
| *Deinococcus, Thermus, Alcaligenes, Cytophaga, Bacteriodes* (all psychrophiles) | South Pole snow[11] |
| *Serratia, Enterobacter, Klebsiella, Yersinia* (all psychrotrophs) | Ellesmere Island ice[21] |
| Viable fungi (*Penicillium, Cladosporium, Ulocladium, Pleurotus,…*) | Greenland ice cores; age ≤ 140,000 yr [54] |
| >57 taxa of eukaryotes (fungi, plants, algae, and protists) | Hans Tausen ice core, northern Greenland[100] |
| *Bacillus* and other soil bacteria | At base of Guliya (Tibet) ice core in 1 My-old ice (J. Reeve, personal comm.) |
| Yeasts, fungi, microalgae, bacteria (including vegetative cells of spore-formers); below 1500 m, only spore- forming bacteria | Vostok ice core[1,2] |
| Non-spore formers (*Pseudomonas…*); spore-formers (mesophiles to psychrophiles); actinomycetes (psychrotolerant) | Vostok ice core[1] |
| Caolobacter, an aquatic oligotroph, probably indigenous to Lake Vostok | Accretion ice at bottom of Vostok core (R. Sambrotto, personal comm.) |
| Aerobic bacteria, mostly psychrotolerant oligotrophic non-sporeformers | Kolyma permafrost[94] |
| 14 diverse genera, dominantly corynebacteria, psychrotrophs, not true psychrophiles † | Kolyma lowland permafrost[87] |
| 11 groups of bacteria including *Proteobacteria and Fibrobacter*; SSU rDNA clones suggest novel genera or families | Kolyma lowland permafrost[104] |
| >30 genera of great diversity, aerobic and anaerobic, including archaea | Kolyma lowland permafrost[34,95] |
| *Bacillus, Arthrobacter, Streptomyces, inter alia* | Antarctic permafrost[95] |
| *Methanococcoides burtonii, Methanogenium frigidum, Halorubrum lacusprofundii* | Psychrophilic archaea in Antarctic lakes[28-30] |

† Shi et al. (87) concluded that the majority of true psychrophiles are found in the ocean. They are rare in Antarctic rocks and soils and permafrost.

Christner *et al.* (16) attempted to cultivate bacteria that they extracted from ice cores from sites including Greenland, Antarctica, Tibet (Guliya glacier), and Bolivia (Sajama glacier). Fewer than 1% could form colonies in a nutrient medium. The success rate was much higher when cultivation took place in a low-nutrient medium over a long time. The authors inferred that cells in low-nutrient media have sufficient time for cell repair before growth begins, whereas cells in rich-nutrient media attempt to grow before they have repaired all the accumulated cell damage. For Sajama ice, they found that the concentration did not decrease with age but may correlate with climatic conditions. For ice from the Guliya glacier, they estimated a concentration of ~$10^4$ cells/ml for the sum of culturable + non-culturable microbes. They found a greater diversity of species in non-polar than in polar ice, which is likely due to proximity to major biological ecosystems.

Carpenter *et al.* (11) found 200 to 5000 cells/ml of bacteria in surface snow and firn from the South Pole. They found that 10 to 20% were members of *Deinococcus*, a genus known to be extremely resistant to ionizing radiation and desiccation. Using tritiated compounds, they found low but non-zero rates of metabolism at ambient temperatures of –12 to -17 C.

Bacterial concentrations were up to $10^6$/ml in Lake Bonney (88,97) and $10^5$ to $4 \times 10^6$/ml in Lake Hoare and other Antarctic lakes (97), usually limited to zones of maximum photosynthesis or concentrated in layers containing sediment particles locally melted due to solar heating.

Zvyagintsev et al. (97) reported concentrations of prokaryotes down to ~60 m depths in Siberian permafrost at –10 to -12 C, ranging from ~300 to ~$10^8$/g, depending on organic content of the soil and percentage of ice in the soil.

Parkes et al. (71) reported the discovery of viable bacterial populations in sea-floor sediment. The concentration decreased logarithmically with depth from as high as $10^{10}$/cm$^3$ at the sea floor to $10^6$/cm$^3$ at a depth 500 m below the sea floor. Karner *et al.* (43) mapped microbes in the North Pacific. Bacteria dominated in water to ~1000 m, below which the concentrations of archaea and bacteria were comparable. Total concentration of the two decreased from ~$3 \times 10^5$/ml at shallow depths to ~7000 just above the sea floor.

Pernthaler *et al.* (72) reported up to $5 \times 10^5$ cells/ml of bacteria and archaea within 8 m of the surface of Lake Gossenköllesee, a high alpine lake.

## A habitat for microbes in solid ice

I recently gave a quantitative argument that, due to their low metabolism, a population of ~10 to ~$10^2$ cells/cm$^3$ can survive for hundreds of thousands of years if they are able to adapt to life in acidic liquid veins in deep ice (78). The conditions would be harsh: no oxygen, no sunlight, high pressure, low temperature, and low nutrient level. Electron microscopic observations (66,102) had shown that in Antarctic ice (but possibly not in Greenland ice (20)) a continuous network of acidic veins exists along triple boundaries of the crystals. Several species of archaea and bacteria are hyperacidophiles, able to live at pH as low as 0 (24,38,49). One species of *Thiobacillus*-like bacterium lives in a Greenland mine (83 N) at pH 0 at a temperature ~ -30 C for the winter months (49).

Micron-sized droplets of acids deposited as aerosols are essentially insoluble in ice crystals. Coarsening of deep ice takes place by migration of grain boundaries, which sweep through and scavenge the droplets. The acid ends up concentrated along veins kept liquid by virtue of its low eutectic temperature. The concentration of acid in veins, $C_{vein}$, is governed by the free-energy requirement that in equilibrium the liquid be on the freezing line. The colder the ice, the more concentrated must be the acid solution in order to keep the liquid from freezing.

The ratio of vein diameter to average grain diameter is proportional to $(C_{bulk}/C_{vein})^{1/2}$, where $C_{bulk}$ is the concentration of acid in meltwater, typically ~1 µM for Antarctic ice. I derived the dependence of acid molarity and of vein diameter on temperature for ice with the impurity composition typical of Antarctica (78). For example, for an average crystal size of 2 cm and an ice temperature of -10 C, corresponding to a location in glacial ice 500 m above Lake Vostok, the vein diameter would be ~7 µm and the acid molarity would be ~2 M.

I showed that microbes in such veins could extract energy from trace quantities of acids brought into the ice as aerosols and could extract carbon and other elements from the constituents of the aerosols. Eqs. [1] and [2] give examples of reactions of aerosol constituents in liquid veins that provide energy, carbon and other elements for biosynthesis when catalyzed by enzymes in microbial membranes:

$$CH_3SO_3^- \rightarrow HS^- + HCO_3^- + H^+ \quad [1]$$
(decomposition of methanosulfonate, yielding 55 kcal/mol at -10 C)

$$4\,HCOO^- + 5\,H^+ + SO_4^{2-} \rightarrow HS^- + 4\,H_2O + 4\,CO_2 \quad [2]$$
(reaction of formic acid + sulfuric acid, yielding 82 kcal/mol at -10 C)

The bulk concentrations of these and a number of other compounds have been measured as a function of depth in melted Antarctic ice (50 and M. Legrand, personal communication). In addition, aerosols undoubtedly deposit polyaromatic hydrocarbons, soluble phosphorus compounds from dead microbes ejected from the ocean surface in storms, and other compounds containing the essential elements for life.

I envisage two ways for bacteria and archaea to establish a habitat in ice veins:

A) <u>Top-down scenario</u>. Wind-blown microbes deposited in snow have a few thousand years in which to prepare for life at low pH; during this time the increasing weight of snow forms firn and, at depths below ~100 m, fully dense ice with a system of interconnected acidic or saline veins. I suspect that in fully densified ice, small microbes will be swept up by migrating grain boundaries, ending up in the veins that comprise triple boundaries. Some microbes can adapt to hostile conditions, either by synthesizing proteins or by evolution. For example, many isolates from glacial ice are found to be highly pigmented, presumably to protect themselves against solar irradiation during atmospheric transport and on the surface of a glacier (16). In culturing microbes in non-enriched, thawed permafrost at 4 C, Vishnivetskaya *et al.* (94) found that most could produce pigments. Because of their very low metabolic activity and low division rate at low temperature, most microbes in ice or permafrost have not evolved much to adapt to their environment; what we see may be the result more of selection than of evolution. At shallow depths, air bubbles form the nodes of the three-dimensional network of liquid veins. Aerobic bacteria, as well as archaea such as *Crenarchaeota* that can tolerate some oxygen, may survive in veins because of access to air.

B) <u>Bottom-up scenario</u>. Microbes in subglacial till, vents, or cracks in bedrock may migrate upward along veins in glacial ice, extracting nutrients and elements for life from redox reactions in the liquid veins. Anaerobic archaea are more likely to be abundant in the bottom-up scenario than in the top-down scenario. Strict anaerobes such as those that inhabit subglacial rock or glacial till may migrate a few meters upward along veins. At depths where all the oxygen from the atmosphere is enclathrated, even methanogens, which cannot tolerate more than ~1 ppm of oxygen, may survive. I note that there is some evidence for adaptation to oxygen by methanogens (105). I also note that even strict anaerobic archaea that die as a result of encountering oxygen can be detected by their autofluorescence when excited at 420 nm, because the $F_{420}$ pigment fluoresces in the oxidized (non-viable) state (61).

# Maximum sustainable microbial population in glacial ice

I estimated (78) that a few cells cm$^{-3}$ could be maintained viable, in a state of "starvation-survival", for ~$4 \times 10^5$ yr if they were able to occupy liquid veins, where the concentration of nutrients is a million times higher than in bulk ice. In a dormant state the population could be far larger, or the time could be far longer. Anderson and Domsch (5) found, for example, that the maintenance requirement of a dormant microbial soil population was lower by a factor ~$10^3$ than for a metabolizing population. Table 3 lists some rates for incorporation of carbon into cells, measured in g carbon (g carbon)$^{-1}$ hr$^{-1}$. Most of the measurements were done in the laboratory under less hostile conditions than those in their natural habitat. The rates should in most cases be regarded as overestimates, since many of the experiments used a rich nutrient in an unconfined medium at room temperature in air, and the population was allowed to grow.

Table 3. Microbial metabolic rates (in g carbon/g carbon/hr) in selected environments

| Site or Organism[ref.] | Method | Temp. | Incorp. Rate (g C/g C hr$^{-1}$) |
|---|---|---|---|
| forest soil, dormant[5] | glucose uptake, $CO_2$ emitted | 28 | $1.7 \times 10^{-4}$ |
| forest soil, dormant[5] | glucose uptake, $CO_2$ emitted | 15 | $1.6 \times 10^{-5}$ |
| Alpine snow[85] | $^{14}$C-leucine uptake rate | 0 | $9 \times 10^{-4}$ |
| S. Pole snow[11] | $^3$H-thymidine; $^3$H-leucine | -17 | detectable |
| *A. globiformis*, starved[14] | C-starved | 25 | ~$5 \times 10^{-5}$ |
| *A. globiformis*, starved[14] | C-starved | 10 | ~$1 \times 10^{-5}$ |
| Ant-300 (*Moritella*)[68] | endogenous metabolic C → $^{14}CO_2$ | 5 | $7 \times 10^{-5}$ |
| deep aquifer[13,74] (sand or clay) | $O_2$ consumption; $CO_2$ production *vs* groundwater age | 20 | $\leq 2 \times 10^{-8}$ |
| seawater[62] | $^{14}CH_4$ oxidation | ? | $(4-80) \times 10^{-5}$ |
| seawater[62] | $^3H_2$ oxidation | ? | $(0.2-11) \times 10^{-5}$ |
| Ross Desert rock[39] | $^{14}$C-bicarb. → $^{14}CO_2$ + lipid C | * | $6 \times 10^{-9}$ |
| Ross Desert rock[39] | $^{14}$C-bicarb. → $^{14}CO_2$ + lipid C | † | $6 \times 10^{-11}$ |
| Frozen tundra[70] | "microbial respiration" | -40 | "detectable" |
| Vostok glacial ice[2] | $^{14}$C-labeled protein hydrolysate | 18 | $8 \times 10^{-5}$ |
| Vostok glacial ice[2] | $^{14}$C-labeled protein hydrolysate | 12 | $\leq 8 \times 10^{-6}$ |

* Optimally sloped rock, above 5 C for 250 h/yr and above 10 C for 90 h/yr
† Horizontally oriented rock, above 5 C only 10 h/yr

Laboratory incubations may greatly overestimate the *in-situ* rates of microbial metabolism. The lowest metabolic rates were for horizontal and inclined rock surfaces containing cryptoendolithic microbial communities in the cold, dry Ross Desert (Antarctica). Next lowest was the rate inferred from geochemical analyses of $O_2$ consumption and $CO_2$ production in groundwater, with age up to 12,000 yr, passing through sediments at depths of hundreds of meters. Radioactive time-course experiments in the laboratory overestimated the metabolic rates of organisms in that groundwater by factors of $10^3$ to $10^6$ relative to the geochemical analysis (74).

At ambient ice temperatures in polar ice (as low as -55 in Antarctic ice and -30 in Arctic ice), all metabolic processes are far slower than in most of the environments where the metabolic rates reported in Table 3 were studied. To estimate the magnitude of the temperature-dependence, I calculated an activation energy based on a best fit to a straight line on an Arrhenius plot. The results for five cases are shown in Table 4. The errors are large, especially for the case of the cryptoendolithic community. In that experiment, Johnston and Vestal (39) collected samples that were growing in the pore spaces of sandstone at an incredibly slow rate and measured the rate of photosynthesis as a function of time, from which they inferred the rate of incorporation of carbon and the turnover time (~20,000 yr!). For a rock surface optimally oriented to receive maximum sunlight, they found that the nanoclimate temperature is above 5 C for ~250 hr/yr and above 10 C for ~90 hr/yr. For a horizontal rock surface the temperature is above 5 C for only 10 hr/yr and never reaches 10 C. They estimated that the annual photosynthesis and carbon incorporation rate in a horizontal rock would be two orders of magnitude less than that in the favorably inclined rock. (See Ross Desert rock, Table 3). Assuming $E_a$ = 262 kJ/mol, and assuming an effective temperature of 0 C, an extrapolation to -30 C would yield a metabolic rate of only ~$10^{-15}$ g C/g C/hr.

Table 4. Activation energies and estimated microbial metabolic rates at given temperatures and sources as derived from data

| conditions [ref.] | data at T | activation $E_a$ | metabolic rate at -30 |
|---|---|---|---|
| Dormant microbes, forest soil[5] | 15 , 28 | 131 kJ/mol | ~7 ×$10^{-10}$ g C/g C/hr |
| Vostok accretion ice[42] | 3 , 23 | 186 kJ/mol | $10^{-4}$ × rate at 3 |
| Vostok glacial ice[2] | 12 , 18 | 264 kJ/mol | ~3 × $10^{-16}$ g C/g C/hr (needs to be higher to fit $N_2O$ data) |
| Permafrost[82] | 5 ,0 ,-1.5 ,-5 , -10 ,-15 ,-20 | 262 kJ/mol | 1/300 × rate at -20 |
| Cryptoendolithic community[39] | seasonally varying | assume 262 kJ/mol | ~$10^{-5}$ × rate at 0 C |

The point of the above exercise, which assumes a temperature-independent activation energy, is to show that at temperatures as low as -30 , metabolic rates become so low that the population of bacteria in liquid veins is very unlikely to be limited by availability of nutrients. It should be no surprise that the majority of cells detected in deep Vostok ice are dormant (2).

In view of the extremely low metabolic rates at low temperatures inferred in the above discussion, I now think a population as large as ~$10^4$ cm$^{-3}$ can be sustained in veins in ice for times greater than $10^5$ years.

## Elemental and isotopic anomalies as evidence for bacteria in glacial ice

Table 5 shows evidence for microbes metabolizing in glacial ice and permafrost. I give a brief discussion of each example and show how experimental data can be used to estimate the metabolic rate of microbes in ice or permafrost as a function of temperature.

Souchez *et al.* (90) reported a dramatic decrease in both $O_2$ and $O_2/N_2$ in the bottom 6 m of the GRIP borehole in contact with subglacial till. They also reported comparably large increases in $CO_2$ (up to 130,000 ppmv), $CH_4$ (up to 6000 ppmv), and $\delta^{18}O$ (from $-38°/_{oo}$, 7 m above the base to $-26°/_{oo}$ at the base). Given a sufficiently large concentration of a substrate such as formate, methanogens can produce both $CO_2$ and $CH_4$ through the reaction

$$4\ HCOOH \rightarrow 2\ H_2O + 3\ CO_2 + CH_4 \qquad [3]$$

Table 5. Excess concentrations of molecules and isotopes: tracers of microbes?

| site, age [ref.] | T | excess gases and isotopes | types |
|---|---|---|---|
| GRIP, 160 kyr [90] | -9 | in lowest 6 m: $1 \times 10^5$ ppmv $CO_2$; 6000 ppmv $CH_4$; ~$10^o/_{oo}$ $\delta^{18}O$ | bacteria + methanogens at shallower depth |
| Dye3, 100 kyr [91] | -13 | $4 \times 10^4$ ppmv $CO_2$ | bacteria |
| Vostok, 138 kyr [92] | -40 | $\delta N_2O \approx 80$ ppbv; $-12^o/_{oo}$ $\delta^{18}O$ of $N_2O$; $15^o/_{oo}$ $\delta^{15}N$ of $N_2O$ | bacteria (nitrifiers?) |
| Sajama, 23 kyr * | -11 | 400 ppbv $\delta CH_4$; $-53^o/_{oo}$ $\delta^{13}CH_4$ | methanogens |
| Miers Valley permafrost, 21 kyr [6] | -17 | $3.6 \times 10^{-4}$ $cm^3/g$ $CH_4$; $-54.8^o/_{oo}$ $\delta^{13}CH_4$ | methanogens |

* Campen and Sowers (personal communication)

Tison *et al*. (93) argued against *in-situ* production of the anomalies in the bottom few m of GRIP ice by microbial metabolism. Nevertheless, I think it would be worthwhile to study microbes in those lowest few meters of GRIP ice to see if their concentrations dramatically increase near the base.

Souchez *et al*. (91) found an excess of nearly 40,000 ppmv of $CO_2$ in the bottom 6 m of Dye 3 (Greenland) silty ice just above subglacial till. The $CO_2$ correlated strongly with the ratio of $N_2/O_2$. $\delta^{18}O$ increased from $-30 \ ^o/_{oo}$ to $-22 \ ^o/_{oo}$ as $CO_2$ increased. The temperature at that depth is -13 C. No studies have been made yet of bacterial concentrations in Dye 3.

Sowers (92) recently reported peaks in the concentrations of $N_2O$ and $\delta^{15}N$ and dips in the values of $\delta^{18}O$ at depths in Vostok ice that correspond to both peaks in bacterial concentration reported by Abyzov *et al*. (2) and peaks in dust concentration (73). These correlations are very striking. Sowers inferred *in-situ* $N_2O$ production by nitrifying bacteria that oxidize ammonia present in the ice. The strongest peak in $N_2O$ is at a depth corresponding to an age of ~138 kyr where the temperature is -40.

K. Campen and T. Sowers (personal communication) have found elevated concentrations of methane and $\delta^{13}CH_4$ at several depths in the Sajama (Bolivia) glacier. At a depth corresponding to an age of ~23,000 yr and T ≈ -11 C, with a total air content of 0.04 $cm^3 g^{-1}$ at Sajama (6543 m above sea level), they found ~$7 \times 10^{-13}$ moles excess $CH_4$ $g^{-1}$ and $\delta^{13}CH_4 \approx -53^o/_{oo}$ in the high-methane regions, compared with $-44^o/_{oo}$ elsewhere. They ascribe both the excess methane and the depleted $\delta^{13}C/^{12}C$ to *in-situ* production by methanogenic archaea.

In their study of microbes in Antarctic permafrost in the Miers Valley with age ~21,000 yr and T ≈ -17 C, Rivkina *et al*. (83) reported finding $3.6 \times 10^{-4}$ $cm^3/g$ methane at a depth of 2 m in lake-bottom sediment. At that same depth they measured $\delta^{13}CH_4 = -54.8^o/_{oo}$, indicating probable biological origin, consistent with the value of $\delta^{13}CH_4$ found by Campen and Sowers. They also detected methanogens and bacteria and used fluorescence to measure a total concentration of up to $10^5$ cells $g^{-1}$. At the workshop, Rivkina showed that $\delta^{13}CH_4$ correlates with $\delta^{13}CO_2$ in such a way as to suggest methanogenesis from $CO_2$ and $H_2$, i.e., isotopically light carbon in methane goes with isotopically heavy carbon in carbon dioxide.

$$4 H_2 + CO_2 \rightarrow CH_4 + 2 H_2O \qquad [4]$$

The data in Tables 3 and 4 provide weak constraints on metabolic rates of microorganisms in ice or permafrost. Because of approximations or extrapolations, they are unreliable, as we will see.

Concentrations of molecules and isotopes of possible biological origin are shown in Table 5, along with the temperature and age of the ice. However, only in the case of Vostok was the concentration of microbes determined at the depth where anomalies in molecular or isotopic composition were reported. For all the other cases, we have to make assumptions. For metabolic production of excess gas, I use the simple relationship

$$Y_k = f_k\, n\, m\, R\, t \qquad [5]$$

where $Y_k$ = biogenic concentration of gas of type $k$, $f_k$ = fraction of reactions going to gas of type $k$, $n$ = concentration of microbes, $m$ = mean mass per microbe, $R$ = relative production rate in g of gas per g of microbes yr$^{-1}$, and $t$ = age of undisturbed ice or permafrost containing the microbes.

For the Sajama glacier Campen and Sowers measured $Y_k$ = 400 ppbv methane = $4 \times 10^{-7}$ cm$^3$ methane per cm$^3$ air; $f_k$ = 1; and $t$ = 23,000 yr. Taking into account their extraction efficiency of 75% and using an air concentration = 0.04 cm$^3$ g$^{-3}$ in ice at the altitude of Sajama (T. Sowers, personal communication) leads to $Y_k \approx 10^{-12}$ moles excess CH$_4$ g$^{-1}$. As eq. [3] shows, formate is a plausible substrate. Although it has not been measured for Sajama ice, the concentration of formate in Vostok ice has been found to be, typically, $\sim 10^{-10}$ mole cm$^{-3}$ of ice. At 1 mole of methane per 4 moles of formate, this concentration would yield up to $2.5 \times 10^{-11}$ mole cm$^{-3}$, a factor 25 more than is needed to produce the observed excess methane. Since $n$ and $m$ for microbes were not measured at the depths where excess gas was observed, I use reasonable guesses: $n = 10^4$ cells cm$^{-3}$ and $m$ = 100 fg cell$^{-1}$. Solving for $R$ leads to a metabolic rate of $\sim 7 \times 10^{-7}$ g methane per g of microbes yr$^{-1}$.

To see if this rate is reasonable, I used the data of Abyzov *et al.* (2) for uptake of $^{14}$C-labeled protein hydrolysate by microbes extracted from Vostok glacial ice. He measured an incorporation rate of $8 \times 10^{-5}$ g C (g C)$^{-1}$ hr$^{-1}$ at 18 and a rate about a factor 10 lower at 12 . Assuming the validity of the Arrhenius temperature dependence leads to an activation energy $E_a$ = 264 kJ/mol. Extrapolating the Arrhenius relation to -11 leads to a predicted rate $R = 4 \times 10^{-6}$ g C (g C)$^{-1}$ yr$^{-1}$, which is within a factor six of the rate I estimated in the previous paragraph. In view of the large uncertainties in values of $n$ and $m$, such agreement is somewhat fortuitous.

In contrast, in the case of N$_2$O in Vostok there is a far worse discrepancy. My calculation, using eq. [5] and an activation energy of 264 kJ/mol, predicts a rate of production of N$_2$O a factor $10^7$ lower than observed by Sowers. Extrapolating Abyzov's rate from 12 down to -40 may account for most of the discrepancy. Others (e.g., 80) have shown that the Arrhenius expression may be fine for chemical rate processes but not for all biological processes. Another possibility is that, despite the impressive correlation with peaks in microbial concentration, the nitrous oxide may not have been produced *in-situ* by nitrifiers.

The point of this discussion is to encourage others to use the method I have outlined to infer microbial metabolic rates at very low temperatures by simultaneously determining concentrations of microbes and concentrations of products of metabolism such as nitrous oxide or methane.

## Three *vs* four physiological states of microbial life

Physiologists and microbiologists have engaged in debate over terminology, especially over whether the term "viable but not culturable" (VBNC) is an oxymoron or not. Recent experiments (41,44,45,59,65,67,96) point toward three categories of physiological state:
• **viability** (i.e. culturability);
• **dormancy** (referred to as "anabiosis" in the Russian literature): a reversible physiological state of low metabolic activity, in which cells can persist for extended periods without division;

• **nonviability** (i.e. death).

Discussion centers on the question of whether there is a fourth state – **viable but non-culturable** (VBNC). It seems to me that culturability is an experimental matter and that failure to culture cells in the laboratory under conditions alien to the environment from which they were removed is not a proof of the existence of the VBNC state.

Recovery from dormancy and ability to divide when conditions improve are now felt to require the presence of either a few viable cells or of a pheromonal substance previously excreted by the viable cells in the medium (41).

Morita proposed replacing the term "maintenance energy" with the term "survival energy", which signifies the minimum level of energy necessary to take care of the destruction of amino acids (racemization) and depurination of the DNA, so that the cell may survive until it meets conditions favorable enough to permit growth and reproduction.

Survival energy results in metabolism that is probably too slow to be measured in the laboratory, but that I believe may be inferred from the rate of production of reactants such as methane, as discussed in the previous section.

## Adaptation to extreme conditions

When subjected to an extremely hostile environment, some microbes continue to metabolize at a greatly reduced rate that depends strongly on temperature and nutrient level; others alter their structure and shut down almost all forms of energy consumption; and still others form spores that enable them to survive without using any energy until conditions improve, after which mechanisms go into action to repair cell damage.

Christner *et al.* (16) found mostly spore-formers in glacial ice, and Abyzov *et al.* (2) reported that most microbes at depths >1500 m are spore-formers. Shi *et al.* (87) found that 30% of permafrost isolates could form spores, a percentage much higher than that usually found in temperate soils (~1%).

Note that it is easier to enter a dormant state than to form spores, which requires more sophisticated structural alterations and takes a long time.

### 1. Starvation

Some microbes react to starvation by developing ultramicrocells (63); some synthesize stress proteins such as the chaperones DnaK and GroEL (48) and Cst and Pex (56); some express starvation genes that confer enhanced general resistance (56); some reduce metabolism and become dormant; some sporulate. *Bacillus*, *Clostridia*, and *Actinobacteria*, all of which are common in ice, are particularly effective in developing a spore with a tough coat. *Dunaliella* can sense high salinity and make up to 30% glycerol that enables it to resist dehydration. Carpenter *et al.* (11) have discovered psychrophilic bacteria that they identified as members of the genus *Deinococcus*. They obtained uptake of tritiated thymidine and leucine, thus showing that in South Pole snow at ambient temperatures of -12 to -17, *Deinococci* are synthesizing DNA and proteins at a low rate.

### 2. Low temperature

Cavicchioli *et al.* (12) and Herbert (36) have summarized some of the responses of microbes to cold shock:

• To counter the loss of membrane fluidity, increase the fraction of polyunsaturated fatty acids.

• To counter lowered rates of enzymatic and transport processes: 1) move, if possible, to an environment richer in organic substrates to compensate for less effective uptake and transport systems; or 2) synthesize a cold enzyme.

• To stabilize nucleic acid secondary structures and their subsequent inhibitory effects on DNA replication, transcription, and translation of mRNA: evolve cold-active proteins so as to produce structurally flexible and catalytically efficient proteins.

• To counter the formation of crystalline ice and its damage to cellular structures, if T goes below the freezing temperature of the cytoplasm: synthesize antifreeze proteins that inhibit formation of ice crystals. Soluble carbohydrates and polyols can also serve as a cryoprotectant.

• Some species naturally have features that confer resistance to freezing and that extend their lifetime under frozen conditions (16). For examples of genera found in glacial ice, see Table 2.

### 3. Low pH

In order to take advantage of nutrients in acidic veins, microbes must be able to tolerate pH as low as 0. For active biosynthesis, cells have to maintain intracellular pH $> \sim 6$. Spore-formers that are unable to maintain that large a pH gradient can differentiate to spores, which in the dormant state resist acids and do not have to maintain the pH gradient. Non-spore-forming acidophiles such as *Dunaliella acidophila* use a combination of methods: a positive inside transmembrane potential; a positive surface charge; a potent plasma membrane $H^+$-ATPase that extrudes protons; a membrane with a very low permeability for protons; a high internal buffering capacity of the cytoplasm arising from the capacity of proteins and other molecules to mop up the incoming protons; and stabilization of proteins and other biomolecules at the external surface of the membrane against acid inactivation (55,75). Because acidophiles are known to include eukaryotes, bacteria, and archaea, one expects to find some members of all three domains in acidic veins.

### 4. Cross-protection

Successful response to starvation enables the cell to resist other stresses such as acid or salinity or high or low temperature (56,64). Either starvation or a low pH environment causes intracellular pH to start to drop, and stimulates the threatened microbe to develop defense mechanisms. For example, Choi *et al.* (15) showed that during starvation, *E. coli* produces a protein, *dps* (DNA-binding protein from starved cells), which protects DNA from various kinds of stress including acid-mediated depurination of DNA. Experiments by Matin (55) showed that metabolically compromised, even nonviable, cells of acidophiles are able to retain a sizable pH gradient for some time after death. A dormant bacterium may be quite dehydrated, so that even if the pH of its cytoplasm drops, acid hydrolysis of chemical bonds will be relatively slow, especially at low temperature.

## "Unfrozen" water in cytoplasm and in permafrost

I now discuss two related problems: 1) How does the cytoplasm avoid freezing if the cell is inside a liquid acidic vein in glacial ice at a temperature as low as –50 ? 2) How do microbes in permafrost at a temperature of -10  to -30  obtain nutrients for metabolism from clay minerals and other fine particles? Elucidating these issues has involved a fascinating interplay among chemists, physicists, biologists, and permafrost specialists.

Earlier I showed that, by virtue of equilibrium thermodynamics, veins at triple boundaries of three grains in polycrystalline ice are kept liquid by a high concentration of ions that suppress the freezing point of water. That is a *volume effect*. The ion concentration in the cytoplasm is too low by itself to keep the cytoplasm from freezing at very low temperature. An additional mechanism is required.

In contrast, the main anti-freeze effect in both the cytoplasm and the water outside the cell membrane is a *surface phenomenon* due to the nature of water itself (mainly its extremely high dipole moment). Inside the cell, water interacts electrically with the charged cell wall and protein network, especially the actin filaments. Outside the cell, water interacts electrically both with the membrane and with fine-grained soil particles. The finer the texture of the sediments, the higher the concentration of viable cells. In loam soil their concentration is two to three orders of magnitude higher than in sands. Furthermore, the concentration of viable cells in permafrost is many orders of magnitude greater than the concentration in pure ice at the same subzero temperature (34,35).

A wealth of experimental measurements add up to a compelling case for "unfrozen" water (also called "structured" water or "layered" water). Experimenters have used various techniques including dielectric relaxation, differential thermal analysis, differential scanning calorimetry, NMR, ultrahigh-frequency dielectric dispersion, and quasi-elastic neutron scattering to infer the fraction of unfrozen water in contact with a surface as a function of temperature below the freezing point of bulk water.

Unfrozen water develops a greater thickness on clay mineral surfaces than on basalt and other rocks (4). They and others showed that in permafrost, a small fraction of the water is in an unfrozen state and envelops organic and mineral particles. The fraction decreases with decreasing temperature, typically following a hyperbolic relation between thickness of unfrozen layer $d$ and $\Delta T \equiv T_m - T$:

$$d = \lambda \, \Delta T^{-n} \qquad [6]$$

(The exponent $n$ is typically ~1/3.) The factor $\lambda$, derived by Wettlaufer *et al.* (99) within their theory of interfacial melting, involves quantities such as latent heat of fusion and free energies of the three interfaces between the solid substance (e.g., clay mineral), ice, and water. Dash *et al.* (22) have shown that the free energy of $H_2O$ and many other solids held at a temperature below the bulk freezing point is lower if there is a thin melted layer on the surface. They call the phenomenon *premelting*. Wei *et al.* (98) have shown that this water layer on ice is present at temperatures above 200K and that the degree of disorder of the water structure increases from 0 at 200K to 100% at 273K.

Using differential scanning calorimetry, McGrath *et al.* (60) found that even at -150 , cells cooled very slowly do not contain intracellular ice, whereas those cooled too rapidly do. The film thickness inferred from the experiments of Anderson (3) ranges from 0.5 to 8 nm depending on temperature and rock composition.

Pollack (76) has discussed in detail the structure and nature of unfrozen water inside cells. The conclusion is that charged surfaces such as occur on proteins attract the oppositely charged ends of water molecules, causing alignment that can extend several layers from the surface. This unfrozen water has different properties from either ice or bulk water, which is why it can be detected experimentally. Because the cytoplasm is crowded with proteins (77), the structured fraction of the cell's water is nearly complete. Typically, each water molecule is no more than about three molecules away from a protein surface. One consequence of the nature of structured water is that the cell does not have to use metabolic energy in order to keep the interior from freezing at very low ambient temperature.

**Potential for life in ice-covered solar system objects**

## 1. Europa – a Jovian moon with an ice-covered ocean

There is growing excitement that Jupiter's moon Europa is a possible location for extraterrestrial biology. Its icy crust may be as thin as ~1 km in certain regions. The Near Infrared Mapping Spectrometer on the Galileo spacecraft (58) has provided evidence for hydrated minerals in the ice, with average composition roughly given by $Na_2CO_3 \cdot 10H_2O + H_2SO_4 \cdot nH_2O + MgSO_4$. Galileo's magnetometer recently measured changes in the magnetic field predicted if a current-carrying outer shell, such as a planet-scale liquid ocean, is present beneath the surface (47). Their evidence that Europa's field varies with time strengthens the argument that a liquid ocean – made conducting by virtue of dissolved salts – exists beneath the present-day surface.

Chyba and Phillips (17) have pointed out that radiolysis by energetic Jovian ions ($H^+$, $O^{6+}$, $S^{6+}$) can produce HCHO and $H_2O_2$, each of which can react further: $2 H_2O_2 \rightarrow 2 H_2O + O_2$, and $HCHO + O_2 \rightarrow H_2O + CO_2$. Irradiation by energetic ions can also produce $SO_2$ and $CO_2$, as well as complex hydrocarbons (18). With recycling of crust into the ocean over $10^7$ yr and a biological turnover time of $10^3$ yr, a steady state number of $3 \times 10^{23}$ cells would result. If they are distributed evenly through an ocean 100 km deep, the concentration would be only 0.1 to 1 cell/cm$^3$, which would seem to make detection impossible. If, however, they were strongly concentrated in nutrient-rich regions near the ice-water interface in the upper 10 to 100 m of the ocean, ice from this layer would yield $10^2$ to $10^3$ cells/cm$^3$. For this range of concentrations, it would seem feasible for an orbiter with an explosive device (Europa Ice Clipper, for example) to collect enough ice to detect microbial life (101).

Tidal flexing results from a Laplacian resonance with 7.2, 3.6, and 1.8 d periods for the three Jovian moons Ganymede, Europa, and Io. This drives tectonics below the ocean floor, keeps Europa's ocean fluid, and probably gives rise to diapirs, which consist of ice warm enough to undergo solid state convection. Ice-penetrating radar on a future Europa orbiter could measure ice thickness and locate diapirs warm enough to contain liquid veins where microbes might exist. The tidal stress generates enough heat and volcanism to melt ice below 10-30 km.

Gaidos *et al.* (32) emphasized that more than only water, hydrocarbons, and other chemicals are needed for life. Without an external energy source such as sunlight or an internal source such as geothermal energy, chemical equilibration would ultimately terminate redox reactions and extinguish any life based on chemical energy. However, given sufficient internal heating (tidal, radiogenic, or chemical) and crustal activity resulting from tidal stress, vertical transport of biological resources to living organisms and waste products from those organisms could be maintained.

Table 6. Some eutectics that could maintain a liquid environment for microbes in a cold planet

| Solute | Eutectic temperature |
|---|---|
| HCl | -88 to -115 C |
| $HNO_3$ | -43 |
| $H_2SO_4$ | -73 |
| Methanosulfonic acid | -75 |
| HCHO | -92 |
| HCOOH | -49 |
| $NaCl \cdot 2H_2O$ | -22 |
| $CaCl_2 \cdot 6H_2O$ | -50 |
| $MgSO_4 \cdot 7H_2O$ | -4 |
| $NH_3 \cdot 2H_2O$ | -93 |
| $CaCl_2 \cdot 6H_2O + MgCl_2 \cdot 12H_2O$ | -55 |

Table 6 lists examples of solutes with low enough eutectic temperatures to provide a habitat for life in diapirs on Europa and other icy planetary bodies.

## 2. Mars – a planet with polar caps of water ice

$CO_2$ ice sublimes at 148K at the Martian surface pressure of 6 mbar. Frozen $CO_2$ alternates between the northern and southern polar caps of Mars with the seasons. In contrast, $H_2O$ ice at the poles is permanent, with a maximum temperature of 205K in summer. With a surface temperature ranging from 154K at the poles to ~218K at the equator and a high incident flux of solar UV, the present Martian surface is hostile to life. Taking into account gradual loss of $H_2O$ from low latitudes and redeposition at the poles (103), it would seem a better bet, if a collector is forced to sample only matter close to the surface, to land in a polar region rather than at a lower latitude. The ice thickness and basal temperature are similar to those for the terrestrial ice caps. Inside the ice, microbes might exist in liquid veins at a temperature above the eutectic for a plausible composition (see Table 6).

Radiogenic heating results in progressively warmer temperatures with depth. Assuming a plausible geothermal heat flux (~30 mW m$^{-2}$), Martian permafrost is expected to vary from 2-3 km at the equator to 6-8 km at the poles. At greater depths, liquid water can exist. Fisk and Giovannoni (26) argued that, at greater depths, conditions similar to those in Earth's deep oceanic subsurface may support life. They present evidence from Martian landers, orbiters, Earth-based observations, and Martian meteorites that the subsurface of Mars meets all the requirements for microbial life: appropriate temperatures, water, carbon, nutrients, metabolic substrates, and a source of energy to maintain chemical disequilibrium. (They recognize, with Gaidos *et al.* (32), that fluid flow or diffusion in the subsurface is needed to maintain metabolism based on chemical energy.) Deep subsurface Martian life may date back several billions of years and consist of anaerobic bacteria or archaea such as methanogens (57). Potential future discovery of isotopically light methane could suggest a biogenic origin.

Comparing the potential biomass of Mars, early Earth, and Europa, Jakosky and Shock (37) estimated that in the last 4 Gyr, as much as 20 g cm$^{-2}$ of biota could have been created on Mars due to hydrothermal circulation and chemical weathering of minerals. This corresponds to ~1 cell cm$^{-3}$ yr$^{-1}$, which is orders of magnitude less than is created on Earth and somewhat greater than can have been created on Europa.

## 3. Ice on Moon and Mercury

Radar observations of Mercury have provided evidence for the existence of substantial deposits of ices in permanently shaded polar regions inside craters with steep walls (9). (This is surprising because Mercury is close to the sun and has an equatorial temperature of ~700K.) A neutron spectrometer on the Lunar Prospector Discovery Mission has provided evidence for water ice in permanently shaded craters on the Moon (25).

Butler (10) calculated that 5 to 15% of all $H_2O$ placed randomly on the surface of Mercury would migrate to stable, permanently shaded polar regions, and that 20 to 50% of $H_2O$ on the moon would migrate to shaded polar regions. Feldman *et al.* (25) estimated that at each lunar pole there are 3 gigatonnes of solid ice down to a depth of 2 m, gardened into a ferro-anorthosite regolith in the top 40 cm. To avoid complete sublimation in 4 Gyr, traps in polar craters on both Mercury and Moon would have to be colder than 112K.

Despite their perennial ice caps, neither the Moon nor Mercury has internal energy sources necessary to generate the chemical disequilibrium to sustain redox reactions necessary for life.

**4. Icy crust and ammonia-water ocean on Titan**

Models of the thermal history of Titan, a moon of Saturn, suggest that an ocean consisting of an ammonia-water solution as much as 200 km deep is presently hidden below a crust of water ice. Fortes (27) argued that all of Titan's current atmospheric $CH_4$ and $N_2$ might have been produced by anaerobic microbes in its ocean during its first $\sim10^8$ years when the ocean was much warmer than 300K. With a present surface temperature of only ~95K, there is little hope of the planned Cassini-Huygens mission observing any chemical signatures attributable to biogenic activity. The best that can be hoped for will be evidence for a subsurface ocean and for fractionation of carbon and nitrogen isotopes in the atmosphere.

Lunine and Stevenson (53) calculated that the eutectic temperature for $H_2O$-$CH_4$ is ~235K at 1 kbar, the estimated pressure at the top of Titan's putative ocean. The possible existence of hotspots, diapirs, liquid veins, and nutrients in the icy crust might lure microbes to migrate toward the surface. Fortes (27) estimated that 8 g cm$^{-2}$ of biota might be produced over 4 Gyr, comparable to that estimated by Jakosky and Shock (37) for Mars. Methanogens near the top of the ocean might be able to deliver sufficient methane to the surface through brittle cracks to buffer the surface atmospheric inventory against loss by photolysis. $N_2$ in the atmosphere might be a biomarker for denitrifying bacteria. Much better measurements of $\delta^{13}C$ of methane and of $\delta^{15}N$ of $N_2$ in Titan's atmosphere would be required in order to provide good evidence for microbial life.

**5. Ice on other satellites of the outer planets**

A spectral absorption feature of $H_2O$ ice, at 1.65 µm, is present in the spectra of all of the icy satellites of Jupiter, Saturn, Uranus, Neptune, and Pluto (19). However, in all but a few cases – Callisto and Ganymede, for example – the surface temperatures are extremely low, and too little is known to draw any conclusions about other conditions necessary for life.

**6. Comets – dirty snowballs**

Comets originate so far from the sun that their temperature is far too low to permit microbial life to arise in their dirty snowball-like nuclei. Their main role in connection with the origin of life on Earth would seem to be as a conveyor of complex organic molecules, synthesized in deep space by ultraviolet irradiation, into Earth's early oceans. Bernstein *et al.* (8) have shown in laboratory experiments that PAHs irradiated with UV lead to synthesis of alcohols, quinones, and ethers. Possibly other prebiotic molecules and amino acids also were brought into the oceans where they served to accelerate chemical reactions key to early life.

## Implications of liquid veins for origin of life at low temperature

High-temperature origin-of-life theories require that the components of the first genetic material be stable. To address this question, Levy and Miller (51) measured the half-lives for the decomposition of the nucleobases and found them all to be short on a geologic time-scale. They concluded that, unless the origin of life took place extremely rapidly, <100 yr, a high-temperature origin of life may be possible but it cannot involve adenine, uracil, guanine, or cytosine. At 0 C, A, U, G, and T appear to be sufficiently stable ($t_{1/2} \geq 10^6$ yr) to be involved in a low-temperature origin of life.

In order to take place in a reasonable time-scale on a cold Earth, a concentrating mechanism is required. Veins provide a mechanism for concentrating prebiotic solutes so as to increase their rates of encounter and of reaction in ice on a cold Earth (6,69). Sanchez et al. (84) pointed out, for example, that the rate of polymerization of HCN, forming part of a plausible prebiotic purine synthesis, is quadratic in cyanide concentration and would be far higher in confined channels than in two or three dimensions.

Stan Miller and co-workers (52) simulated prebiotic synthesis on Europa and other ice-covered planets and satellites by studying the synthesis of organic compounds from dilute solutions of ammonia + cyanide kept frozen for 25 years at -78 C (eutectic T ≈ -100 C). They found the reaction rate to be higher at -78 than in a liquid at 25 C! After 25 years the ice had a brown color, and adenine, guanine, and amino acids had been produced in substantial yields. I am convinced that, had they looked in a microscope before melting the ice, they would have found that the brown color was confined to liquid veins in the ice.

## Concluding Remarks

1. As yet there is no experimental evidence to show whether or not microbes take advantage of the habitat I have proposed. No one has yet looked for microbes in glacial ice before melting it. Junge *et al.* (40) have introduced a dye into brine channels in sea ice and have used epifluorescence to show that microbes do move in those channels. However, this is a special case: the channels are usually open to the ocean on which the ice floats, so that microbes readily move in and out. Furthermore, the channels are in many cases huge ("pockets"). Their very nice study is an encouraging first step toward finding whether micron-sized acidic or saline veins in glacial ice also provide a habitat for life.

2. At this workshop Bay (7) discussed two methods we are developing for studying microbes *in-situ* in solid ice:

 • a biospectrologging tool that can be lowered into a fluid-filled borehole in glacial ice in order to make *in-situ* measurements of fluorescence by microbes in each of the three states (viable, dormant, or dead);

 • epifluorescence microscopy in a cold box, with and without a live/dead stain, to scan liquid veins in solid ice in order to search for microbes and attempt to assay the fraction that are in each of the three states.

3. A proof that microbes live in veins in ice and a measurement of the fraction of viable or dormant *vs* dead ones in veins relative to those in the solid phase would have a number of interesting consequences:

 • It would show that certain microbes can simultaneously tolerate six or seven adverse environmental parameters: low temperature, high pressure, low pH or high osmolarity, low nutrient concentration, low oxygen concentration, and absence of sunlight.

 • Such microbes may be even more exotic than those that may exist in subglacial lakes isolated from the Antarctic surface for many millions of years. By contrast, in subglacial lakes only the oxygen concentration and pressure, not all six parameters, are inhospitable to microbial life.

 • Microbes able to survive for geologic time at pH 0 and at a temperature as low as -50 in terrestrial ice would serve as a proxy for microbial life in diapirs in Europan ice.

4. A quantitative relationship between microbial mass concentration, concentration of metabolic product, and age of ice or permafrost at different temperatures would tell us the metabolic rate of the microbial population *in-situ*, a result that cannot be obtained in laboratory experiments at low temperatures. To be convincing, the concentrations of microbes and metabolic products must correlate with depth in the ice or permafrost.

5. It should be no surprise that dormant cells in deep glacial ice are difficult to cultivate in a laboratory at 1 atmosphere, at room temperature, in air, etc. To consider just one parameter – pressure – I point out that piezophiles (also known as barophiles) adapted to a pressure of 300 bars in the Pacific Ocean have a greatly reduced growth rate at 1 bar. Microbes near the bottom of the Vostok core (~3600 m depth) will have been at a similar pressure and will not grow well at 1 bar.

6. The nature of unfrozen water both inside cells and in permafrost, and the layer thickness on surfaces as a function of temperature, are on a rather firm experimental basis, and there is a good theoretical foundation for unfrozen water on surfaces outside cells.

**Acknowledgments**